\documentclass[twocolumn,showpacs]{revtex4}
\usepackage{amsfonts}
\usepackage{amsmath}
\usepackage{graphicx}
\begin{document}
\draft
\title{AC conductivity of a niobium thin film in a swept
magnetic field}
\author{M.I. Tsindlekht$^1$, V.M. Genkin$^1$, $\check{\text{S}}$. Gazi$^2$, and $\check{\text{S}}$. Chromik$^2$ }
\affiliation{$^1$The Racah Institute of Physics, The
Hebrew University of Jerusalem, 91904 Jerusalem,
Israel}

\affiliation{$^2$The Institute of Electrical Engineering SAS, D$\acute{u}$bravsk$\acute{a}$ cesta 9, 84104  Bratislava, Slovakia}

\begin{abstract}

We report the results of the measurement the ac conductivity of a Nb superconducting thin film in a swept dc magnetic field.  In the mixed state the swept dc field creates vortices at the film surface which pass through the film and form the observed ac conductivity. Vortex rate generation does not depend on the value of the dc field and there is a large plateau-like region of dc magnetic fields where the dissipation is approximately constant. A proposed phenomenological model describes quite well the main features of the ac response in these fields including its dependency on the  sweep rate, ac amplitude, frequency, and value of the second and third harmonics.

\end{abstract}
\date{\today}
\maketitle
\section{Introduction}

It is known that the ac response of type II bulk superconductors in slow ramped dc fields differs qualitatively from the ac response in constant dc fields~\cite{STR2}. Increased ac losses in the mixed state~\cite{MAX} and the second harmonic generation~\cite{CAMP} were observed in swept dc fields. Sweeping of a dc field induces dc current in the sample that changes both components of the ac response, while the dc current, which also could exist in the sample due to pinning forces, can change only the penetration depth~\cite{PROZO}, but does not increase the ac absorption. It was found that both real, $\chi_1^{\prime}$, and imaginary, $\chi_1^{\prime\prime}$, components of the susceptibility depend on the dimensionless parameter $q=\dot{H}_0/\omega h_0$, where $h_0$ is an ac amplitude, $\omega$  is the frequency, $H_0$ is the external dc field, and $\dot{H}_0$  is the sweep rate~\cite{SCH}. Several models were discussed in the literature. A bulk dissipation mechanism associated with flux-flow or flux-creep, was considered and was found inadequate \cite{MAX}, because the loss component $\chi_1^{\prime\prime}$ actually does not depend on the dc field for $H_{c1}<H_0<H_{c2}$. The phenomenological switching model~\cite{MAX, SCH,FINK2,CAMP} supposes that if $q<1$ the instantaneous time rate of the field changes its sign for a fraction of each ac period. During this interval the vortices become pinned and the sample is lossless~\cite{PARK}. During the remainder of the period the loss mechanism should operate. The sample is switched back and forth from the dissipative state to the non-dissipative state and the resistivity reaches some average value. The difficulty with this model is the lack of losses in the mixed state in constant dc fields ~\cite{STR2,ROLL,KL,SHT}, and it is not clear which resistivity has to be averaged.

The effect of the swept dc field on the ac response was discovered a long time ago, but up to now there are few measurements of the ac conductivity in this case.
Our experiments with bulk samples~\cite{GENKIN} showed that in the swept dc field the conductivity of the sample could not be characterized by a single value and one has to consider the conductivity that depends on the distance from the sample surface. Interpretation of these experimental data is complicated by the inhomogeneity of the ac electric field. The amplitude of the ac electric field is decreased inside the sample and the response of a bulk sample is then some average value. To overcome these difficulties we have to simplify the problem. Experiments with thin films can provide this simplification.
For a thin film, which forms the wall of a hollow cylinder with actually arbitrary shape of a cross section, in longitudinal ac magnetic field the electric field in the film is homogeneous with accuracy $\approx \text{d}/L$, d is a film thickness and $L$ is the some macroscopic length. This permits us to introduce the averaged over the thickness conductivity  $\sigma(\omega)=\sigma_1(\omega)+i\sigma_2(\omega)$ of the film and to measure this quantity. For a long sample the screening currents in the walls circulate only in the plane which is perpendicular to the rectangular like cylinder axis and, consequently, both the electric field and current are constant along this contour.

In this paper we report the experimental results measurements of the ac conductivity of a Nb polycrystalline thin film in a swept dc field applied parallel to the surface. The film was deposited on the four sides of the parallelepiped sapphire substrate. We show that in the mixed state the dc field generates vortices at the surface, which cross the film and form the observed ac response.
Numerical simulations qualitatively describe the experimental data below $H_{c2}$, and make clearer the physical picture. Due to the sweeping of dc magnetic field the vortices in the sample are always at the threshold of the depinning. Only during part of an ac period vortices are depinned and provide the penetration of the ac field through the film. These results are reminiscent the assumption of the switching model~\cite{MAX, FINK2}, but our simulations indicate that the pinned-depinned transition does not take place when the time derivative of the external magnetic field changes its sign. We also find  that increasing the ac amplitude only slightly increases the vortex flow across the film. The ratio of the vortex number, which crosses the film, to the ac amplitude characterizes the observed magnetic susceptibility,  increasing of $h_0$ increases shielding and decreases losses as is observed in the experiment. For $H_0>H_{c2}$ the physical picture is not clear yet and more research is needed.

\section{Experimental details}

\begin{figure}
\begin{center}
\leavevmode
\includegraphics[width=0.9\linewidth]{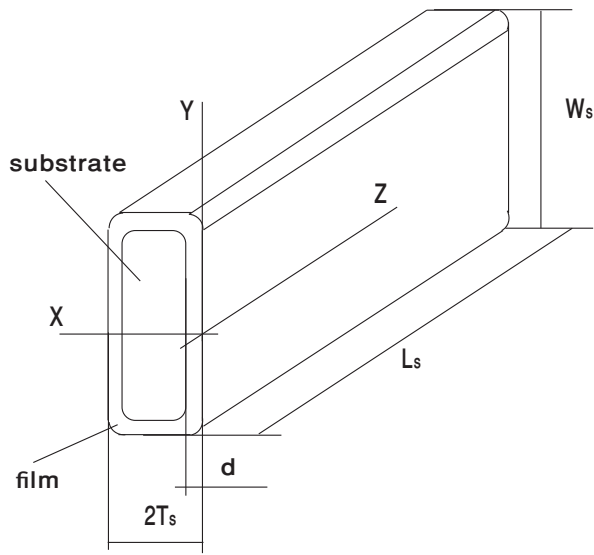}

\caption{Sketch of the S2 sample. Here $\text{L}_s = 15$ mm, $\text{W}_s=3$ mm, and $2\text{T}_s=1.5$ mm are the substrate length, width and thickness, respectively. Film thickness is d=200 nm. Both dc and ac fields are parallel to $Z$-axis. All dimensions are not in the scale.}
\label{f-0a}
\end{center}
\end{figure}

 The Nb films were deposited by dc magnetron sputtering at room temperature. Two different Nb thin samples have been prepared and measured. Sample S1 is the 600 nm film that was deposited on the one side of the substrate. Sample S2 is a 200 nm film which was deposited on the four sides of the substrate. The sizes of the sapphire substrate with rounded corners (radius 0.2 mm) are 1.5 by 3 by 15 mm. Actually we formed a thin-walled hollow superconducting cylinder with rectangular cross section.
The sketch of the S2 sample is shown in Fig.~\ref{f-0a}.

 DC magnetic properties were measured using a standard SQUID magnetometer. The ac response was measured by the pick-up coil method. The sample was inserted into one of a balanced pair of coils, and the unbalanced signal was measured by a lock-in amplifier. A "home-made" measurement cell of the experimental setup was adapted to a commercial SQUID magnetometer. The block-diagram of the experimental setup has been published elsewhere~\cite{LEV2}. The magnetic susceptibilities of the sample at frequencies 293 and 1465 Hz and $h_0$  from 0.04 to 1.2 Oe were measured in two modes. The first one is a point-by-point mode when, during the measurement the dc field was kept constant, and the second one is a swept field mode in which the dc field was ramped at a given rate. The external ac and dc fields were parallel to the films surface. For the measurements in a swept field, the standard power supply of the SQUID magnetometer
solenoid was replaced by an external Oxford Instruments superconducting magnet power supply.

In zero dc field and for low temperatures the superconducting film completely shields the small external ac field if $\lambda^2/L\text{d}<<1$, where $\lambda$ is the London penetration depth~\cite{KITTEL}. The observed susceptibility of the sample in this case equals $-1/4\pi$. This allowed us to obtain the ac susceptibility in absolute units for any field and temperature. Measurements were performed at two temperatures, 7 and 8 K.

\section{Experimental results}

\begin{figure}
\begin{center}
\leavevmode
\includegraphics[width=0.9\linewidth]{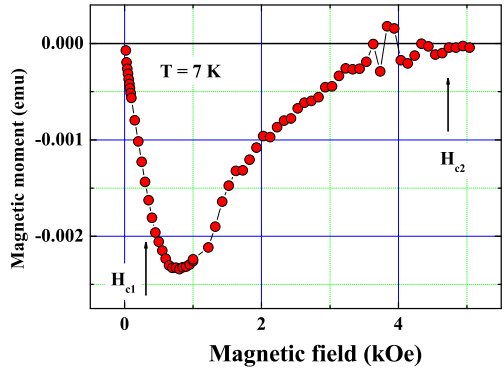}

\caption{(Color online)  ZFC magnetization curve of the Nb film deposited on the one side of a substrate, S1 sample,  at T = 7 K. }
\label{f-0b}
\end{center}
\end{figure}

Fig.~\ref{f-0b} shows the isothermal zero-field cooled (ZFC) magnetization curve of S1 sample at 7 K.
These data permitted us to estimate $H_{c1}$ as 300 Oe and $H_{c2}$ as 4.7 kOe, and the correlation length and London penetration depth are 25 and 70 nm respectively~\cite{PG}.  The residual resistance ratio, $R_{300 K}/R_{10 K}$, of this film is $\approx 4$ and its critical temperature T$_c\approx 8.5$ K.

\begin{figure}
\begin{center}
\leavevmode
\includegraphics[width=0.9\linewidth]{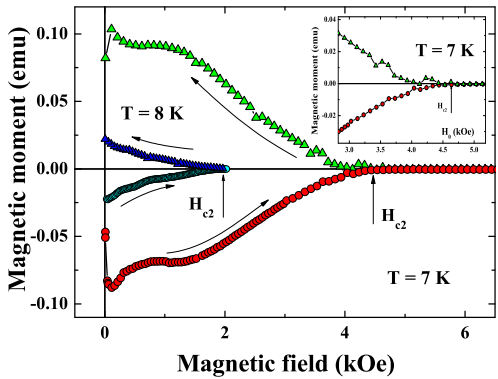}

\caption{(Color online) ZFC magnetization curve of the S2 sample with Nb film deposited on the four sides of a substrate at T = 7 and 8 K. Inset: magnetization curve near $H_{c2}$ at T = 7 K.}
\label{f-1a}
\end{center}
\end{figure}

The magnetization curves of the S2 sample at 7 and 8 K are shown in Fig.~\ref{f-1a}.
The observed magnetic moment of the S2 sample is formed by the magnetization of the film itself and by the current circulated in the film around the substrate. The volume of the film is  $V_{f}=2.7\times 10^{-5}$ cm$^3$ while the substrate volume is $V_s=6.7\times 10^{-2}$ cm$^{3}$. Since $V_s/V_f\approx 10^{3}$, the contribution of the film itself is negligible. Fig.~\ref{f-0b} shows that actually the magnetic moment of the film itself does not exceed 2.5$\times10^{-3}$ emu.

Fig.~\ref{f-1a} shows that at 7 K the penetration of the dc field through the film begins at $H_0< 300$ Oe possibly due to defects in the film.
The dc magnetic moment of both samples actually disappears at 7 K for $H_{0}> 4.6$ kOe. Since no other transition is observed for $H_0>4.6$ kOe (see inset to Fig.~\ref{f-1a}), we conclude that both samples have the same $H_{c2}\approx 4.6$ kOe at 7 K.

\begin{figure}
\begin{center}
\leavevmode
\includegraphics[width=0.9\linewidth]{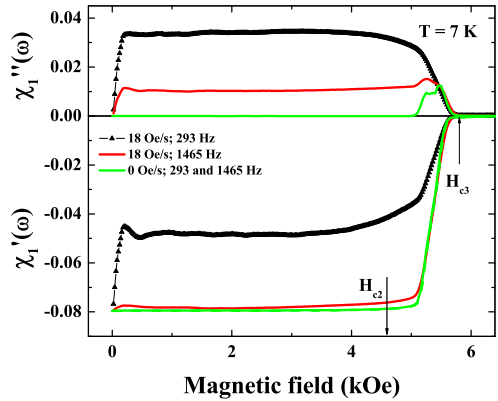}
\includegraphics[width=0.9\linewidth]{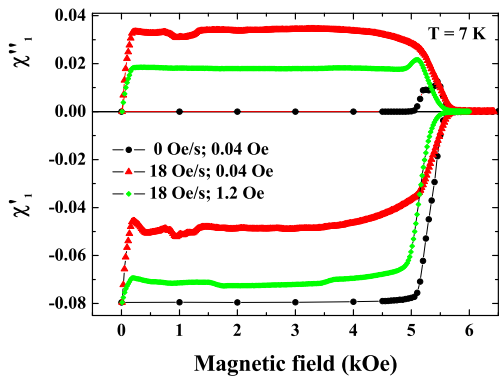}

\caption{(Color online)  Field dependencies of the real and imaginary components of the magnetic susceptibility of sample S2. Upper panel: frequencies 293 and 1465 Hz, and amplitude 0.04 Oe. Lower panel: amplitudes 0.04 and 1.2 Oe, and a frequency of 293 Hz. Measurements were carried out in point-by-point mode (sweep rate 0 Oe/s) and in a swept field mode with a sweep of rate 18 Oe/s. }
\label{f-2a}
\end{center}
\end{figure}

Fourier analysis of the ac magnetization yields an expression of the form $$m(t)=h_0\sum_n\chi_n\exp(-in\omega t).$$ Susceptibilities $\chi_1$, $\chi_2$ and $\chi_3$ were measured. Upper panel of Fig.~\ref{f-2a} shows $\chi_1$ of the S2 sample as a function of the dc magnetic field at T = 7 K, for ac amplitude 0.04 Oe, frequencies 293 and 1465 Hz, measured in constant dc field (point-by-point mode) and in a swept dc field (swept mode) with a rate of 18 Oe/s.
Point-by-point data (zero sweep rate) do not show any difference of $\chi_1$ for frequencies 293 and 1465 Hz, while in a swept dc field $\chi_1$ depends considerably on the frequency. In the swept field $\chi_1^{\prime\prime}$ arises at low magnetic fields. Similar to the bulk superconductors ~\cite{STR2,MAX,GENKIN}, there is a large plateau-like region of magnetic fields where $\chi_1^{\prime\prime}$ is approximately constant while in constant field dissipation and incomplete shielding are observed only for $H_0> H_{c2}=4.6$ kOe, i.e. in the area of surface superconductivity. The difference between the point-by-point and swept field data becomes smaller as the dc field approaches $H_{c3}$. As well as in bulk Nb~\cite{GENKIN} amplitude of excitation affects the ac response. Fig.~\ref{f-2a} (low panel) shows $\chi_1$ field dependence of the S2 sample at two amplitudes of excitation 0.04 and 1.2 Oe for sweep rate 18 Oe/s. Increase of the ac amplitude leads to decrease the losses and increase screening in plateau-like region.

Experimental data can be presented in terms of the average conductivity of the film.
Since the ac susceptibility is caused by the total current in the film $\chi_1\text{S}h_0 =j_s(\omega)\text{Sd}/c$,  where $\text{S}=\text{W}_s\times 2 \text{T}_s$  is the area of the cross-section perpendicular to the field direction (see Fig.~\ref{f-0a}) and $j_s(\omega)$ is the average current density in the film at the fundamental frequency. We neglected the demagnetizing factor, because it is small ($\approx 0.036$). From Maxwell's equation $\text{curl}\overrightarrow{E}=i\omega \overrightarrow{B}/c$ we obtain the electric field in the film as $e_0=i\omega \text{S}(1+4\pi \chi_1)h_0/c\text{L}$, where $\text{L}=2\text{W}_s+4\text{T}_s$. The average conductivity $\sigma(\omega)=j_s/e_0$ is
\begin{equation}\label{Eq2a}
    \sigma(\omega)=\sigma_1+i\sigma_2=-\sigma_0\frac{i\chi_1\omega_0}{[1+4\pi\chi_1]\omega},
\end{equation}
where $\sigma_0=c^2\text{L}/\omega_0\text{Sd}\approx 1.6\times 10^{14}~(\text{Ohm}\times\text{cm})^{-1}$ for S2 sample and $\omega_0/2\pi = 1$ Hz. The skin depth corresponding to this conductivity  is $ 8\times 10^{-3}$ cm at frequency 1 Hz.  It is worth noting that the imaginary part of the conductivity of any superconductor in the Meissner state is $\sigma^{\prime\prime}_\text{L}=c^2/4\pi\lambda^2\omega_0\approx 3\times 10^{17}$ $(\text{Ohm}\times\text{cm})^{-1}$, for the London penetration depth $\lambda = 70$  nm, while the normal conductivity of a pure single Nb crystal is approximately $\sigma_n\approx 10^6-10^7 $ $( \text{Ohm}\times\text{cm})^{-1}$ \cite{GENKIN}.

\begin{figure}
\begin{center}
\leavevmode
\includegraphics[width=0.9\linewidth]{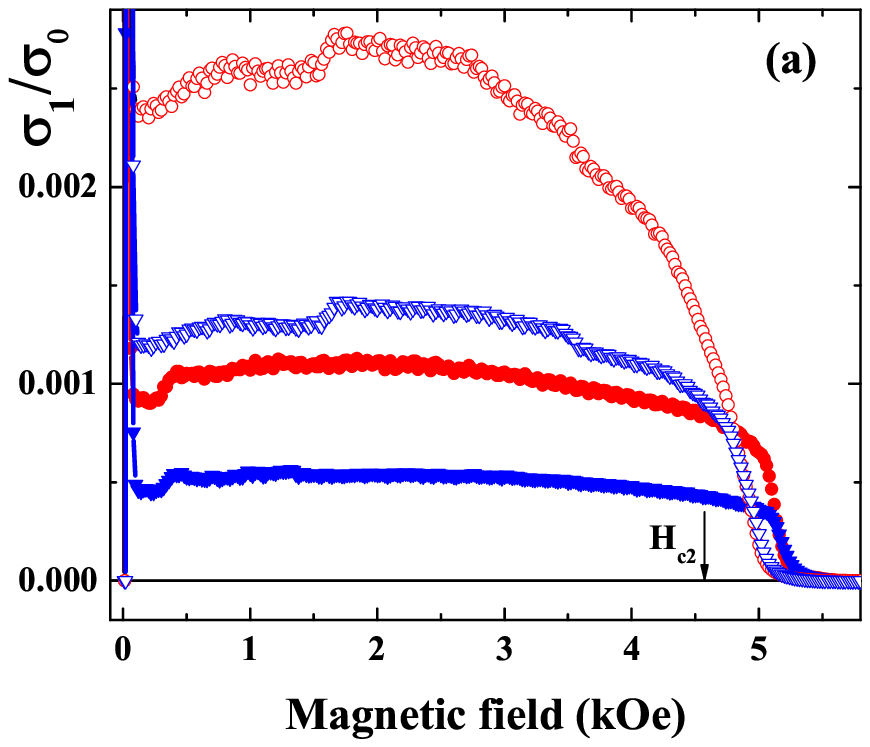}
\includegraphics[width=0.9\linewidth]{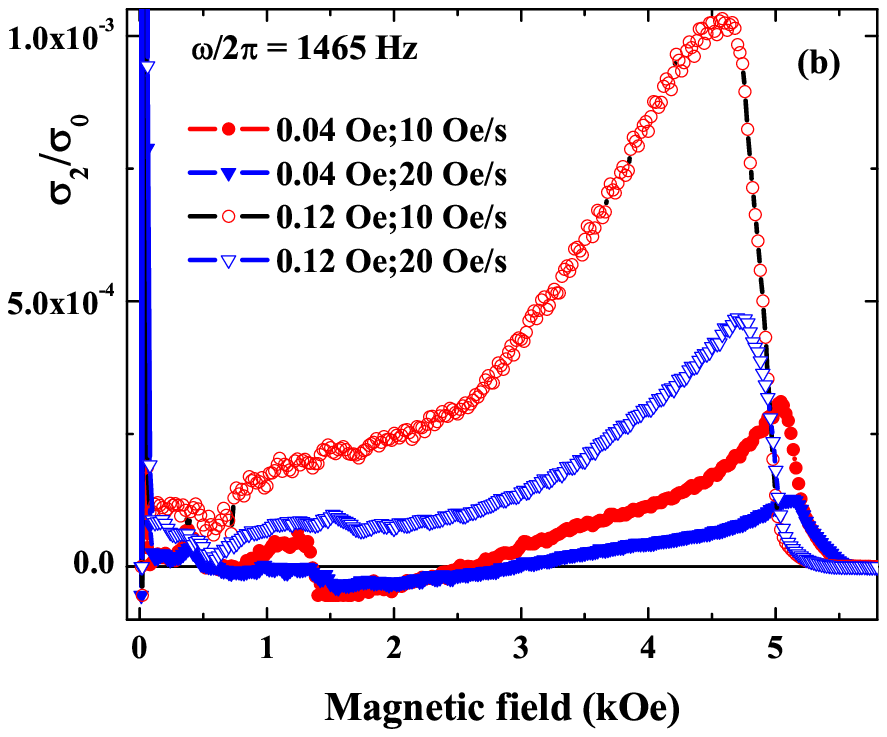}

\caption{(Color online)   Real (panel $a$) and imaginary (panel $b$) components of the conductivity versus magnetic field at frequency 1465 Hz and T = 7 K. }
\label{f-3a}
\end{center}
\end{figure}
\begin{figure}
\begin{center}
\leavevmode
\includegraphics[width=0.9\linewidth]{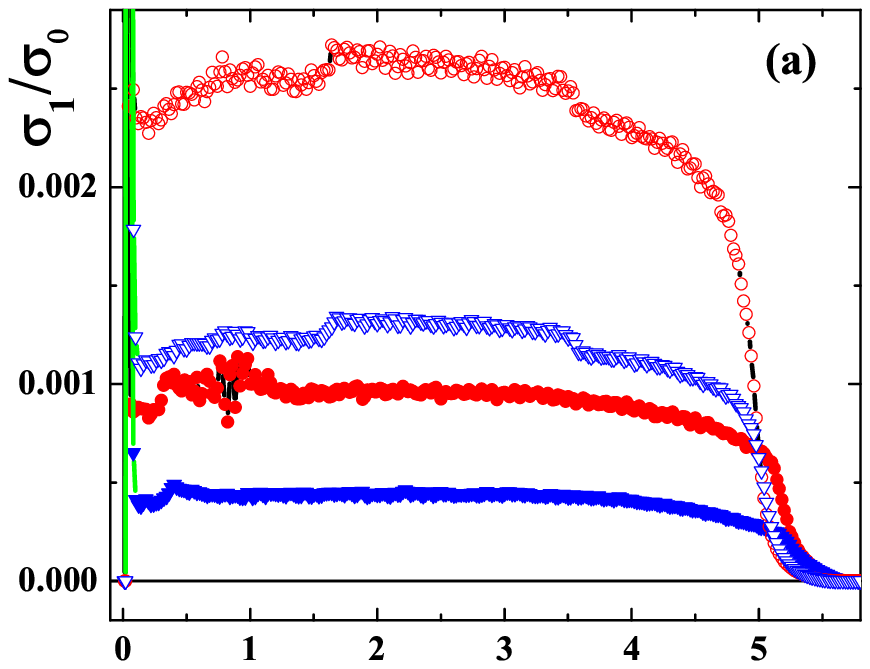}
\includegraphics[width=0.9\linewidth]{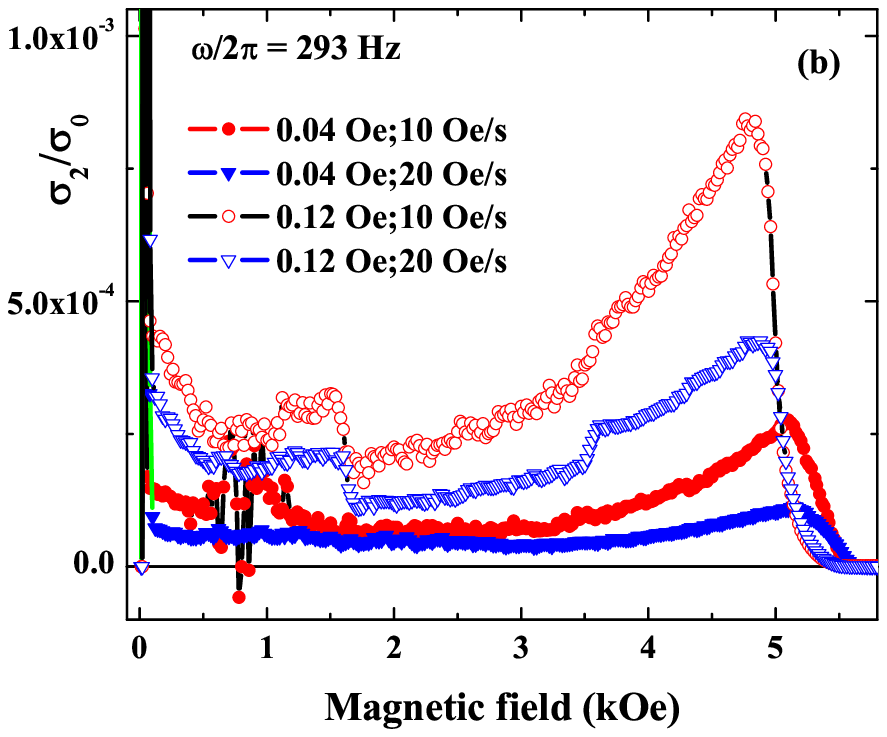}

\caption{(Color online)  Real (panel ($a$) and imaginary (panel $b$) components of the conductivity versus magnetic field at frequency 293 Hz and T = 7 K. }
\label{f-4a}
\end{center}
\end{figure}
The conductivity $\sigma(\omega)$ is extracted from the obtained data of $\chi_1$ by using Eq.~(\ref{Eq2a}). Figs.~\ref{f-3a} and \ref{f-4a} demonstrate the field dependencies of the real and imaginary part of the conductivity at different frequencies, ac amplitudes, and sweep rates. Both real and imaginary components of $\sigma$ are decreasing with the sweep rate, but increasing with the ac amplitude. The frequency dependence of $\sigma$ is comparatively weak. On the other hand, $\chi_1^{\prime\prime}$ exhibits approximately $1/\omega$  frequency dispersion, Fig.~\ref{f-2a}.
$\chi_1\approx -1/4\pi$  at 1465 Hz in magnetic fields smaller than 3 kOe, and in this field region using Eq.~(\ref{Eq2a}) requires more precise measurements. This is the reason for the negative value of $\sigma_2$ in Fig.~\ref{f-3a} and \ref{f-4a}.
The measured conductivity is considerably larger than the conductivity of high quality single crystal in the normal state, thus the Bardeen-Stephen formula $\sigma=\sigma_nH_{c2}/H_0$ cannot describe the experimental data adequately. However, the conductivity of the film induced by swept field is approximately two orders of magnitudes smaller than the bulk conductivity of a niobium sample in a mixed state in a swept field, see Fig. 7 at Ref.~\cite{GENKIN}.

\section{Discussion}

\begin{figure}
\begin{center}
\leavevmode
\includegraphics[width=0.9\linewidth]{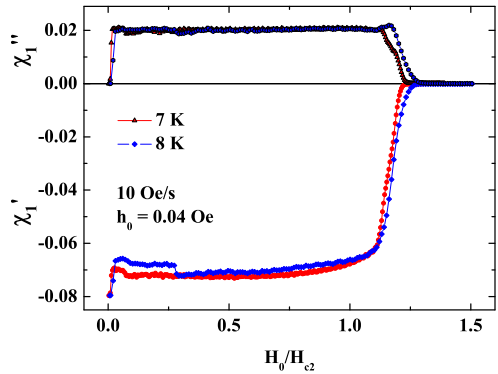}

\caption{(Color online)  Field dependence of $\chi_1$ in a swept field with rate 10 Oe/s, $\omega/2\pi =293$ Hz, and $h_0=0.04$ Oe at T = 7 and 8 K.}
\label{f-7a}
\end{center}
\end{figure}

We will discuss the obtained experimental results on the basis of the following model.
The substrate dimensions satisfy the following inequality $\text{T}_s\ll \text{W}_s \ll \text{L}_s$. The external magnetic fields are directed along the $Z$ axis, Fig.~\ref{f-0a}. The thickness of the deposited superconducting film is $\text{d}\ll \text{T}_s$. We describe the vortices in the film in term of the vortex density $\rho(x)$, neglecting the vortex lattice structure. In this approximation both the vortex density and the magnetic field depend only on the coordinate $x$, \textit{X} -axis is normal to the film surface, see Fig.~\ref{f-0a}. This approach requires an averaging over distances larger than the vortex spacing. For a thin film this condition is not well satisfied. However, the averaging in plane parallel to the film surface completely flattens out the periodic dependence in the normal to the film direction if the film has roughness larger than the period of the vortex lattice. In this case the roughness of the polycrystalline film could expand the applicability of approach used for thin films. The continuity equation for the $\rho(x)$ is
\begin{equation}\label{Eq3a}
\frac{\partial\rho}{\partial t}+\frac{\partial}{\partial x}\biggl[V\times\rho-D\times\frac{\partial \rho}{\partial x}\biggr] = 0,
\end{equation}
where $V$ is the mean vortex velocity and $D$  is the diffusion coefficient.
The diffusion constant in Eq.~(\ref{Eq3a}) is small because we observed the difference between applied and internal (in the substrate) dc fields, Fig.~\ref{f-1a}. For the large diffusion constant the magnetic moment S2 sample should be the same order as magnetic moment of S1 sample, see Fig.~\ref{f-0b}. Diffusion is a temperature activated process and therefore its contribution have to depend on the temperature. In Fig.~\ref{f-7a} we show the experimental $\chi_1$  as a function of the reduced dc field $H_0/H_{c2}$ for T = 7 K ($H_{c2}= 4.6$ kOe) and 8 K ($H_{c2}= 1.9$ kOe), while frequency, ac amplitude, and sweep rate) were the same. These curves are practically identical and we may neglect the diffusion term in Eq.~(\ref{Eq3a}).

At low frequencies the vortex velocity depends on the current density and the pinning force~\cite{COCL}. In this case the simplest approximation for the vortex velocity is $V=0$ if $|\partial H/\partial x|$ is smaller than maximum value defined by pinning forces $(|\partial H/\partial x|)_{pin}\equiv F_p$. Here $H$ is the magnetic field in the film.
The equation for $H$ is:
\begin{equation}\label{Eq3b}
\frac{\partial^2 H}{\partial x^2}+\frac{\rho \phi_0-H}{\lambda_{eff}^2}  = 0,
\end{equation}
where $\lambda_{eff}$ is the effective penetration depth, and $\phi_0$ is the flux quantum. Taking into account the elasticity of the vortex lattice $\lambda_{eff}^2=\lambda^2+\lambda_C^2$, where $\lambda$, $\lambda_C$ are the London and the Campbell penetration depths, respectively~\cite{COCL,CAMPB}.
Eq.~(\ref{Eq3a}) describes the vortex motion in the film, while $\lambda_C$ in Eq.~(\ref{Eq3b}) takes into account the displacement of the vortex lattice from its equilibrium position. These  displacements also  provide ac losses in the constant dc field, but they are small at low amplitude of excitation and we do not see them. So, in our approximation we could assume that the vortex velocity equals zero if the maximal value of the pinning force is larger than the Lorentz force.
The curl of the magnetic field in the film is:
\begin{equation}\label{Eq4a}
\text{curl}\overrightarrow{H} = (\phi_0\nabla \theta/2\pi-\overrightarrow{A})/\lambda^2,
\end{equation}
where $\overrightarrow{A }$ is the vector-potential, $\text{curl}\overrightarrow{A }=\overrightarrow{H}$,  and $\theta $ is the phase of the order parameter. Integration over the contour which encircles the substrate ($\text{T}_s\ll\text{W}_s $) yields
\begin{equation}\label{Eq5a}
    2\lambda^2dH(\text{d})/dx=\phi_0\Phi/2\pi\text{W}_s-\text{T}_sH(\text{d}),
\end{equation}
where $H(\text{d})$ is the magnetic field at the internal surface of the film and $\Phi$ is the change in phase after a complete circulation around the substrate. $\Phi$ obeys the equation:
 \begin{equation}\label{Eq6a}
    d\Phi/dt=4\pi\text{W}_s\phi_0J_v,
\end{equation}
where $J_v$ is the vortex flow into the substrate.
Equations (\ref{Eq5a}, \ref{Eq6a}) are boundary conditions for magnetic field $H$ at $x=\text{d}$.
Let $H(t,x)=H_{dc}(t,x)+h_{ac}(t,x)$, where $H_{dc}(t,x)$ is the slowly swept dc field and $h_{ac}(t,x)$ is the ac field. If the vortex flow, $V\times\rho$, is zero due to pinning then $d\Phi/d t=0$ and $$\biggl[\frac{2\lambda^2}{\text{T}_s}\biggr]\times dh_{ac}(\text{d})/dx+h_{ac}(\text{d})=0.$$
Estimating $dh_{ac}(\text{d})/dx\approx h_{ac}(\text{d})/\text{d}$ we obtain the dimensionless parameter $p=\lambda^2/\text{dT}_s$ in this expression. If $p\ll 1$ then $h_{ac}(\text{d})= 0$. This means that the ac magnetic field in this limit is completely shielded by the film. This effect was discussed in~\cite{KITTEL}.
Experiment shows that for the sweep field, when $H_{dc}<H_{c2}$, the ac field penetrates into the substrate, while in a constant dc field ac field penetrates at $H_{dc}>H_{c2}$, Fig.~\ref{f-2a}.

Equations (\ref{Eq3a},~\ref{Eq3b}) with proper boundary conditions, Eqs.~(\ref{Eq5a}, \ref{Eq6a}), were solved numerically. A grid with 200 points along the $X$-axis was taken and the space derivatives were approximated by finite differences. The time evolution of the obtained ordinary differential equations was found by the forward Euler method.
The vortex velocity $V$ as a function of $\partial H/\partial x$ is approximated by the following expression
\begin{eqnarray}\label{Eq12a}
V=0, ~\text{if}~ |\partial H/\partial x|<F_p ~\text{and}~~~ \nonumber\\
V=-A_0\times{\partial H/\partial x}\times{\frac{z^2}{1+z^2}},~z\equiv A_1\times{(|\partial H/\partial x|-F_p)}~\\ \text{for}~ |\partial H/\partial x|>F_p.~\nonumber
\end{eqnarray}
Here $A_0$ and $A_1$ are phenomenological parameters. The function $z^2/(1+z^2)$ smooths $V(\partial H/\partial x)$ dependence near $|\partial H/\partial x|-F_p=0$.
Taking into account that in constant dc field ac field does not penetrate to the substrate we accept that $\lambda_{eff}/\text{d}\approx 0.3$.
Susceptibilities  $\chi_n$  could be found through the Fourier components of the magnetic field at the inner surface of the film, i. e. at $x=\text{d}$, when at $x=0$ the magnetic field is $H_e=H_{0}(t)+h_0\sin(\omega t)$.
The parameters $A_0$ and $A_1$ were found by fitting the calculated $\chi_1$ to the experimental $\chi_1$ for 3 kOe, 293 Hz, ac amplitude 0.04 Oe and sweep rate of 10 Oe/s.  $A_0$ and $A_1$ parameters remain unchanged for other values of the magnetic field for the plateau region. Obtained values of $A_0$ and $A_1$ were used to calculate  $\chi_1$, $|\chi_2|$, and $|\chi_3|$ for other ac amplitudes, frequencies, and sweep rates.

\begin{figure}
\begin{center}
\leavevmode
\includegraphics[width=0.9\linewidth]{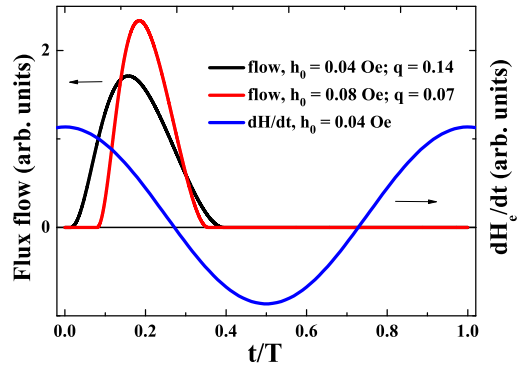}

\caption{(Color online) Flux flow at the boundary film-substrate and the derivative of applied field $dH_e/dt$  as a function of time, for ac amplitude $h_0=0.04,~ 0.08$ Oe, sweep rate 10 Oe/s, and frequency 293 Hz. \text{T} is a period of an ac field.}
\label{f-8a}
\end{center}
\end{figure}

Calculated vortex flow, normalized by the $\omega\times d$ and with $\rho=1$, at the boundary film-substrate ($x=\text{d}$) as a function of time for two ac amplitudes 0.04 and 0.08 Oe, sweep rate 10 Oe/s and frequency 293 Hz is shown in Fig.~\ref{f-8a}. The function $dH_e/dt=\dot{H}_0+\omega h_0\cos(\omega t)$ is also plotted in Fig.~\ref{f-8a}.
The flux flow is nonzero only during the part of the ac period as was assumed by the switching model~\cite{MAX,FINK2,CAMP}. However, the time when the flux flow becomes zero does not coincide with the time when the time derivative of the applied magnetic field changes its sign. Using data from Fig. \ref{f-8a} we can see that
the maximal vortex velocity is the order of $10^{-2}-10^{-3}$ cm/s. Rough estimation of the scale vortex velocity in our experimental arrangement could be obtained from such considerations. Approximately the rates of increasing both applied and internal (in the substrate) dc fields are the same, and  $dN/dt=\dot{H}_0\times 2T_sW_s/\phi_0$, where N is the number of flux quanta in the substrate. On the other hand, this quantity could be written as $dN/dt=2H_0V W_s/\phi_0$, and we obtain for the vortex velocity the simple expression $V=T_s \dot{H}_0/H_0$. For $T_s\approx1$ mm, sweep rate 20 Oe/s and dc field 2 kOe we obtain a vortex velocity $V\approx 10^{-3}$ cm/s, that agrees well with numerical results. The vortex velocity measured in a magnetic field applied perpendicular to thin Nb film is a few orders of magnitude lager~\cite{SCS}. The geometry of the experiment could be a reason for the difference between our result and the result obtained in Ref.~\cite{SCS}.

\begin{figure}
\begin{center}
\leavevmode
\includegraphics[width=0.9\linewidth]{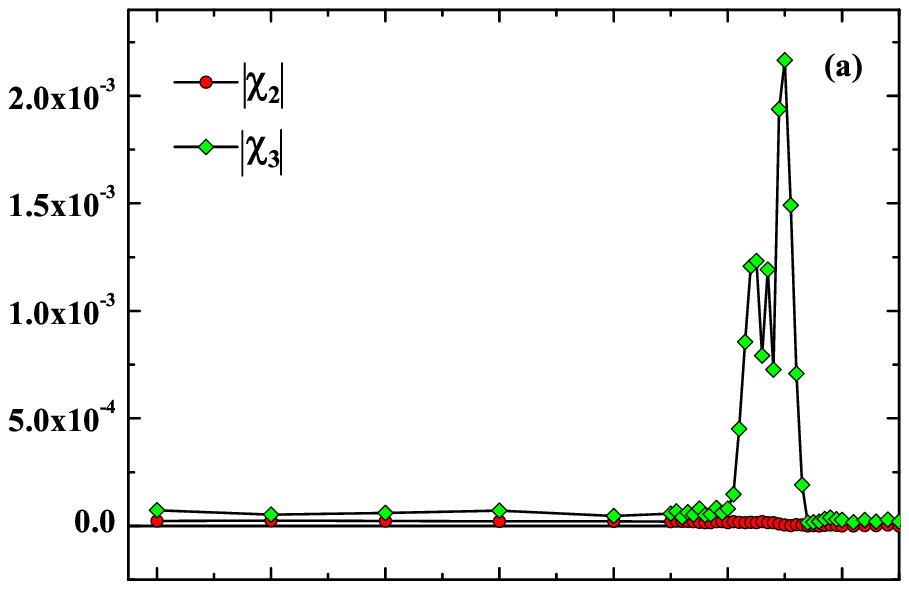}
\includegraphics[width=0.9\linewidth]{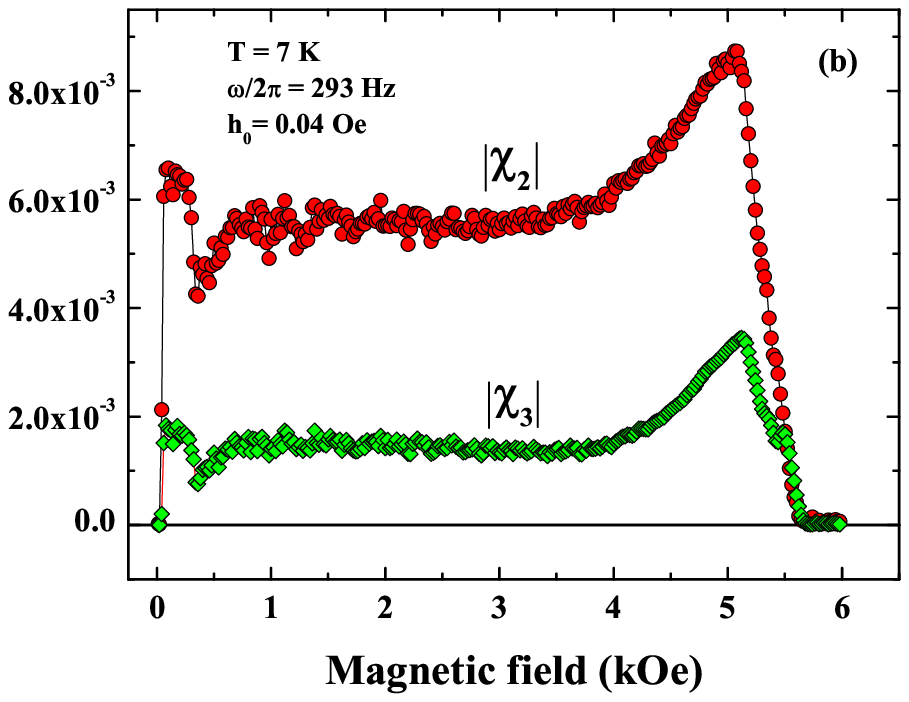}

\caption{(Color online) Field dependence of $\chi_2 $ and $\chi_3$ in point-by-point mode (panel $a$) and in a swept magnetic field with rate 10 Oe/s (panel $b$).}
\label{f-9a}
\end{center}
\end{figure}

Fig.~\ref{f-9a} demonstrates $|\chi_2|$ and $|\chi_3|$ as a function of the dc field at 293 Hz, $h_0 =0.04$ Oe for point-by-point and swept field modes with the sweep rate 10 Oe/s. In the mixed state in constant dc fields $|\chi_2|$ and $|\chi_3|$ are zero due to the pinning, while in the swept field the vortices in the sample are at the threshold of the depinning, and both the second and third harmonics are generated. The flux flow waveform explains the appearance of the second and third harmonics in a swept dc field in the mixed state.

\begin{figure}
\begin{center}
\leavevmode
\includegraphics[width=0.9\linewidth]{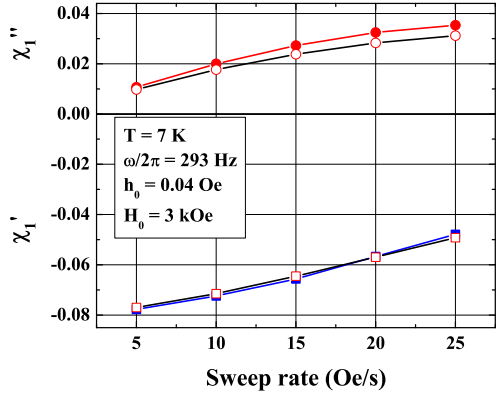}

\caption{(Color online) Sweep rate dependence of $\chi_1$ at T = 7 K, $h_0=0.04$ Oe and $H_0=3$ kOe. Open symbols - theoretical data, closed symbols - experimental data.}
\label{f-10a}
\end{center}
\end{figure}

Experimental and theoretical data for $\chi_1$, $|\chi_2|$ and $|\chi_3|$ at two frequencies, two amplitudes for $H_0=3$ kOe and sweep rate 10 Oe/s are shown in Table~\ref{T1}.
\begin{table}   \centering   \caption{Experimental and theoretical values of $\chi_1$, $|\chi_2|$ and  $|\chi_3|$ at T = 7 K, $H_0=3$ kOe and sweep rate 10 Oe/s. }\label{T1}
\begin{tabular}{|c|l|c|c|c|c|c|} \hline
$h_0$ (Oe)&&Frequency (Hz)&$\chi_1^{\prime}$&$\chi_1^{\prime\prime}$&$|\chi_2|$&$|\chi_3|$\\ \hline
0.04&theory&293&-0.071&0.017&0.007&0.003\\ \hline
0.04&exp.&293&-0.072&0.020&0.005&0.001\\ \hline
0.08&theory&293&-0.076&0.010&0.004&0.002\\ \hline
0.08&exp.&293&-0.077&0.011&0.003&0.001\\ \hline

0.04&theory&1465&-0.079&0.004&0.002&0.0006\\ \hline
0.04&exp.&1465&-0.079&0.004&0.0009&0.0002\\ \hline
0.08&theory&1465&-0.080&0.002&0.0009&0.0005\\ \hline
0.08&exp.&1465&-0.079&0.002&0.0007&0.0002\\ \hline
\end{tabular}
\end{table}
The theoretical model describes the experimental data in the plateau-like region reasonably well.
Increasing the amplitude by two times ($0.04-0.08$ Oe) does not significantly increase the vortex flow (Fig.~\ref{f-8a}) while $\chi_1^{\prime\prime}$ decreases by a factor of two, as it was found in experiment.
There is a good agreement between the calculated and experimental values of $\chi_1$ as a function of the sweep rate. For frequency 293 Hz, $h_0$ = 0.04 Oe, and dc field 3 kOe this dependence is shown in Fig.~\ref{f-10a}.
The frequency dispersion of $\chi_1$ is also described well by the theoretical model, Table~\ref{T1}, while for harmonic generation the relation between the model and experiment is not so good.

\section{Conclusions}

We have investigated the low frequency ac response of a thin niobium film in point-by-point and a swept dc field modes. We obtained the low-frequency conductivity of the film in swept dc fields using experimental data for the ac susceptibility of a thin-walled hollow superconducting cylinder with rectangular cross section. It was found that the conductivity in a mixed state depends on the excitation amplitude, frequency and sweep rate. The Bardeen-Stephen formula cannot describe adequately the experimental data for conductivity. A model that deals with the ac response of the film in the mixed state has been proposed. This model is based on the continuum approximation. The model assumes that in a swept field vortices are at the threshold of the depinning. In this case a superimposed weak ac field yields the possibility of the vortex moving during a part of the ac period only. As a result, an ac field penetrates through the film into the substrate and the losses appear. The calculated waveform of the vortex flow at the film-substrate boundary explained the appearance in a mixed state the second and third harmonics in swept dc fields. We have to note that the accepted model is rather rough and, in spite of this, theoretical data are in a good agreement with experiment in the plateau-like region. In the point-by-point mode the losses, ac field penetration through the film and the third harmonic generation were observed only for dc fields larger than $H_{c2}$. The physical picture of the ac response in the surface superconducting state, $H_0>H_{c2}$, for bulk and thin film samples is not yet clear and additional studies are needed.

\section{Acknowledgments}

The authors are deeply thankful to J.R. Clem, I. Felner and G.I. Leviev for valuable discussions.  We thank J.R. Clem who kindly drew our attention to paper~\cite{KL}.
This work was supported by the Klatchky foundation for superconductivity. We also acknowledge a grant from the VEGA agency for financial support for projects Nos. 2/0173/13.

\end{document}